\documentclass[12pt]{paper}
\usepackage[T1]{fontenc}
\usepackage[a4paper]{geometry}
\usepackage{amsbsy}
\usepackage{amstext}
\usepackage{cancel}
\usepackage{esint}

\usepackage[normalem]{ulem}
\usepackage[dvips]{color}
\usepackage{booktabs}
\usepackage{threeparttable}
\usepackage{graphicx}
\usepackage{CJK}

\makeatletter
\usepackage{amsfonts}

\makeatother
\renewcommand{\sout}{\bgroup \color[rgb]{1,0,0}\ULdepth=-.5ex \ULset}

\begin{document}

\title{Proof of Atiyah-Singer Index Theorem by Canonical Quantum Mechanics}

\author{Zixian Zhou, Xiuqing Duan, Kai-Jia Sun}
\maketitle
\begin{abstract}
We show that the Atiyah-Singer index theorem of Dirac operator can
be directly proved  in the canonical formulation of quantum mechanics,
without using the path-integral technique. This proof takes advantage
of an algebraic isomorphism between Clifford algebra and exterior
algebra in small $\tau$ (high temperature) limit, together with simple
properties of quantum mechanics of harmonic oscillator. Compared to
the proof given by heat kernel, we try to prove this theorem more
quantum mechanically. 
\end{abstract}

\section{Introduction\label{sec:Introduction}}

Atiyah-Singer index theorem \cite{Ati68}, which states that the analytical
index of an elliptic complex equals to the topological index of the
corresponding fiber bundle, connects analysis to topology in an insightful
way. In particular, the index theorem of Dirac operator \cite{Fri00}
is extremely important in mathematics and physics. It not only unifies
 the Gauss-Bonnet-Chern theorem \cite{Che45}, Hirzebruch-Riemann-Roch
theorem \cite{Hir54}, and Hirzebruch signature theorem \cite{Hir78}
through vector bundle isomorphism, but also plays a central role
in some novel topology-induced physical phenomena, like chiral magnetic
effects in heavy-ion collisions \cite{Fuk08}, and condensed matter
physics \cite{Li16}, as well as for the analysis of zero energy modes
of the graphene sheet \cite{Pac07,Die16} et al.

In the past several decades, mathematicians have explored various
methods based on topology and analysis including cobordism \cite{Ati63,Pal65},
embendding \cite{Ati68}, heat kernel \cite{Ati63-1} to prove the
 Atiyah-Singer index theorem. Moreover, there are also more physical
proofs based on path integral or probability \cite{Par82} in quantum
mechanics (QM) \cite{Gau83,Fri85} with an underlined key element
called supersymmetry (SUSY).  As we know, QM can be equivalently
formulated in different frameworks: the canonical formulation in Hilbert
space, Feynman path integral, and the QM in phase space \cite{Wig32,Moy49,Gro46,Sza03}
et al. In the SUSY-QM proof by path integral, one needs to take great
care of the measure of both Grassmannian and ordinary variables in
infinite dimensional integral. In the present work, we try to show
that the index theorem of Dirac operator can be concisely proved in
the canonical formulation of QM.

In our proof, the Laplace operator in Hodge's theory \cite{Nak03}
is treated as the Hamiltonian, and the analytical index (also Witten
index) is expressed as an integration on the whole manifold. By establishing an algebraic isomorphism between the
exterior algebra and the Clifford algebra \cite{Law89} in small $\tau$
limit, we can quickly simplify the integrand to be a topological characteristic.
This step plays a similar role as the localization technique \cite{Gau83,Sch97}
in the path integral of the SUSY-QM proof, and is also essential for
the heat kernel's proof \cite{Ezr85}. Moreover, the $\hat{A}$ characteristic
can be straightforwardly obtained by the skill of canonical transformation
in QM. It is found to be closely related to the quantum harmonic oscillators.
Throughout the derivation, the SUSY is naturally encoded in Hodge's
theory and Clifford algebra. Therefore, our proof can be considered
as parallel to the previous SUSY-QM proofs by path integral.

\section{Preliminary and notations}

\subsection{Analytical index of Dirac operator}

We first briefly introduce a $2n$-dimension (closed) spin
manifold $M$ ($n\in N$) and notations. Its Riemann metric takes
$\delta_{\mu\nu}e^{\mu}\otimes e^{\nu}$ ($\mu$ runs from 1 to $2n$
in Einstein summation convention) in the local unitary tangent frame
$e_{\mu}\left(x\right)\in\Gamma\left(TM\right)$ and cotangent frame
$e^{\mu}\left(x\right)\in\Omega^{1}\left(M\right)$. The coefficients
of Levi-Civita connection are given by $\Gamma_{\mu\nu}^{\lambda}\left(x\right)$
which induce the curvature 2-form $R=\frac{1}{2}e^{\mu}\land e^{\nu}R_{\beta\mu\nu}^{\alpha}\left(x\right)e_{\alpha}\otimes e^{\beta}$.
The tangent bundle $TM$ and cotangent bundle $T^{\ast}M$ further
induce the Clifford bundles $C\ell\left(M\right)$ and $C\ell^{\ast}\left(M\right)$,
with anti-commutation relation $\left\{ \tilde{e}_{\mu}\left(x\right),\tilde{e}_{\nu}\left(x\right)\right\} =-2\delta_{\mu\nu}$
and $\left\{ \tilde{e}^{\mu}\left(x\right),\tilde{e}^{\nu}\left(x\right)\right\} =-2\delta^{\mu\nu}$
respectively \cite{Law89}. Here a negative sign is added in front
of metric $\delta_{\mu\nu}$ so that the Clifford algebra naturally
represents a super complex number. (Compared with the  convention
$\left\{ \gamma_{\mu},\gamma_{\nu}\right\} =2\delta_{\mu\nu}$ in physics, we
have $\tilde{e}_{\mu}=i\gamma_{\mu}$.) Note there is an natural linear
isomorphism $\varphi:\land\left(TM\right)\rightarrow C\ell\left(M\right)$
that $\varphi\left(e_{\mu_{1}}\land\cdots\land e_{\mu_{r}}\right)=\tilde{e}_{\mu_{1}}\cdots\tilde{e}_{\mu_{r}}$and
$\varphi\left(e^{\mu_{1}}\land\cdots\land e^{\mu_{r}}\right)=\tilde{e}^{\mu_{1}}\cdots\tilde{e}^{\mu_{r}}$.

The spin manifold $M$ induces a spinor bundle $\Delta\left(M\right)=\Delta^{+}\left(M\right)\oplus\Delta^{-}\left(M\right)$,
where the section of $\Delta^{\pm}\left(M\right)$ represents a right/left-handed
fermion field. The connection form and curvature form of the spinor
bundle are given by $\tilde{\Gamma}=\frac{1}{4}e^{\alpha}\Gamma_{\alpha\mu}^{\nu}\tilde{e}^{\mu}\tilde{e}_{\nu}$
and $\tilde{R}=\frac{1}{8}e^{\mu}\land e^{\nu}R_{\beta\mu\nu}^{\alpha}\left(x\right)\tilde{e}_{\alpha}\tilde{e}^{\beta}$,
respectively. Moreover, the fermion field may have interaction through
a gauge field, with gauge potential $\omega$ and strength $\Omega$.
In mathematics, this is formulated by a twisted product between the
spinor bundle $\Delta\left(M\right)$ and a Hermitian vector bundle
$\pi:E\rightarrow M$, with connection 1-form $\omega=e^{\alpha}\omega_{\alpha}$
and curvature 2-form $\Omega=\frac{1}{2}e^{\mu}\land e^{\nu}\Omega_{\mu\nu}$
of $E$. Therefore the spinor bundle becomes the twisted Dirac vector
bundle $\Delta\left(M\right)\otimes E$ and its connection is given
by $e^{\alpha}\otimes D_{\alpha}$ with 
\begin{equation}
D_{\alpha}=\partial_{\alpha}+\frac{1}{4}\Gamma_{\alpha\mu}^{\nu}\tilde{e}^{\mu}\tilde{e}_{\nu}+\omega_{\alpha}\label{eq:conn}
\end{equation}
as a differential operator on the space of section $\mathcal{H}=\Gamma\left(\Delta\left(M\right)\otimes E\right)$.
Here $\partial_{\alpha}$ denotes the directional derivative along
$e_{\alpha}$.

The twisted Dirac operator for the fermion field, a self adjoint first-order
elliptic differential operator, is defined by $\cancel{D}=\tilde{e}^{\alpha}D_{\alpha}:\mathcal{H}\rightarrow\mathcal{H}$.
In physics it is usually expressed as $i\gamma^{\alpha}D_{\alpha}$
and arises in Dirac equation. The Dirac operator maps the right/left-handed
fermion field to the left/right-handed one $\cancel{D}^{\pm}=\tilde{e}^{\alpha}D_{\alpha}:\mathcal{H}^{\pm}\rightarrow\mathcal{H}^{\mp}$
with subspace $\mathcal{H}^{\pm}=\Gamma\left(\Delta^{\pm}\left(M\right)\otimes E\right)$.
The analytical index of $\cancel{D}^{+}$ is defined by 
\begin{equation}
\textrm{ind}\cancel{D}^{+}=\textrm{dim}\left(\textrm{ker}\cancel{D}^{+}\right)-\textrm{dim}\left(\textrm{coker}\cancel{D}^{+}\right).
\end{equation}
Since $\cancel{D}^{\pm}$ is self-adjoint Fredholm operator, the analytical
index also equals to $\textrm{ind}\cancel{D}^{+}=\textrm{dim}\left(\textrm{ker}\cancel{D}^{+}\right)-\textrm{dim}\left(\textrm{ker}\cancel{D}^{-}\right)$,
which can be regarded as the difference of the degeneracy of the ground
states between the right-hand and left-hand fermions, in other word,
the asymmetry of the chirality of ground states. Then Atiyah-Singer
index theorem for twisted Dirac operator claims that its analytical
index equals to its topological index,
\begin{equation}
\text{ind}\cancel{D}^{+}=\int_{M}\hat{A}\left(TM\right)\land\text{ch}\left(E\right),
\end{equation}
in which $\hat{A}\left(TM\right)$ is the $\hat{A}$ genus of the
manifold $M$ and $\text{ch}\left(E\right)$ is the Chern character
of vector bundle $E$.

\subsection{Witten index}

The coboundary (nilpotent) operator $\cancel{D}^{+}$
in the analytical index can be replaced by Laplacian,
\begin{equation}
\textrm{ind}\cancel{D}^{+}=\textrm{dim}\left(\textrm{ker}\Delta^{+}\right)-\textrm{dim}\left(\textrm{ker}\Delta^{-}\right),\label{eq:Hodge}
\end{equation}
in which $\Delta^{\pm}=\cancel{D}^{2}:\mathcal{H}^{\pm}\rightarrow\mathcal{H}^{\pm}$
is the Laplace operator on the right-handed fermion field (positive
spinor) and left-handed fermion field (negative spinor), respectively.
(Please notice $\Delta$ denotes the Laplace operator while $\Delta\left(M\right)$
denotes the spinor bundle.) From Hodge's theory we learn: (1) the
eigen value of the Laplacian $\Delta=\cancel{D}^{2}:\mathcal{H}\rightarrow\mathcal{H}$
is non-negative; (2) $\textrm{ker}\Delta=\textrm{ker}\cancel{D}^{+}\cap\textrm{ker}\cancel{D}^{-}$;
(3) $\left[\cancel{D}^{\pm},\Delta\right]=0$ . We can regard the
self-adjoint elliptic operator $H=\frac{1}{2}\Delta$ as a quantum
mechanical Hamiltonian operator which determines the statistics of
the quantum state and $\mathcal{H}$ as the corresponding Hilbert
space. We immediately obtain an identity $\left[\cancel{D}^{\pm},H\right]=0$
which is known as supersymmetry, a symmetry between fermion and boson
in quantum mechanics. Therefore, the SUSY is naturally encoded in
the Hodge's theory on the spinor bundle. Assuming $\mathcal{H}_{\lambda}^{\pm}=\left\{ \psi|H\psi=\lambda\psi,\psi\in\mathcal{H}^{\pm}\right\} $
denotes the right/left-handed eigen subspace of $H$ with eigen value
$\lambda$, then the isomorphic relation $\cancel{D}^{\pm}\mathcal{H}_{\lambda}^{\pm}\cong\mathcal{H}_{\lambda}^{\mp}$
can be easily established for arbitrary positive energy eigenvalue
$\lambda>0$ by the properties (1)\textendash (3), which leads to
$\textrm{dim}\mathcal{H}_{\lambda}^{+}=\textrm{dim}\mathcal{H}_{\lambda}^{-}$
for $\lambda>0$. Evidently, the ground-state subspaces with different
chirality are not isomorphic to each other, and Eq. (\ref{eq:Hodge})
actually tells $\textrm{dim}\mathcal{H}_{0}^{+}-\textrm{dim}\mathcal{H}_{0}^{-}=\textrm{ind}\cancel{D}^{+}$. 

Then the analytical index can be reduced to Witten index \cite{Wit82}.
Following Ref. \cite{Wit82}, the fermion number operator $\left(-1\right)^{F}:\psi^{\pm}\mapsto\pm\psi^{\pm},\forall\psi^{\pm}\in\mathcal{H}^{\pm}$
is defined to separate\sout{s} different chirality,
\begin{equation}
\left(-1\right)^{F}=i^{n}\tilde{e}^{1}\cdots\tilde{e}^{2n}=\varphi\left(\ast i^{n}\right),\label{eq:chir}
\end{equation}
with $\ast$ being the Hodge star. It is also called the complex volume
element in complexified Clifford algebra $\mathbb{C}\ell$ and the
chirality operator in quantum field theory. Thus, the analytical index
in Eq. (\ref{eq:Hodge}) is reduced to Witten index 
\begin{equation}
\textrm{ind}\cancel{D}^{+}=\text{tr}_{\mathcal{H}}\left(-1\right)^{F}e^{-\tau H},\label{eq:Witten}
\end{equation}
in which the contributions of the positive energy states to the index
cancel completely to each other but only those of the ground states
remain. Here $\tau$ can be an arbitrary number and we choose $\tau>0$
for simplicity. Moreover, the expression of the Hamiltonian in Eq.
(\ref{eq:Witten}) is given by the Weitzenb\"{o}ck identity of Laplacian
\cite{Gri78},
\begin{equation}
H=\frac{1}{2}\left[\sum_{\alpha=1}^{2n}\left(-D_{\alpha}^{2}+\Gamma_{\alpha\alpha}^{\nu}D_{\nu}\right)+\frac{\mathcal{R}}{4}+\tilde{\Omega}\right],\label{eq:Hamil}
\end{equation}
with $\tilde{\Omega}=\varphi\left(\Omega\right)=\frac{1}{2}\tilde{e}^{\mu}\tilde{e}^{\nu}\Omega_{\mu\nu}$
a curvature 2-form and $\mathcal{R}$ the Ricci scalar. 

\section{Proof of index theorem}

\subsection{QM on curved manifold}

The key framework of our proof is the Dirac notation of QM on curved
manifold . Each point $x\in M$ correspond to a ket $\left|x\right\rangle $
which forms the basis of Hilbert space $L^{2}\left(M\right)$. Provided
coordinate $\boldsymbol{q}=\left(q^{1},\cdots,q^{2n}\right):U\rightarrow\mathbb{R}^{2n}$
on open set $U\subset M$, the ket is defined by unitary relation
\begin{equation}
\left\langle x|x'\right\rangle =\frac{1}{\sqrt{g\left(x\right)}}\delta\left[\boldsymbol{q}\left(x\right)-\boldsymbol{q}\left(x'\right)\right],\label{eq:unitary}
\end{equation}
in which $g=\det\left(g_{\mu\nu}\right)$ is the determinant of the
metric $g_{\mu\nu}\left(x\right)\text{d}q^{\mu}\otimes\text{d}q^{\nu}$.
Since the right-hand-side of Eq. (\ref{eq:unitary}) is a geometric
invariant, the ket defined in $L^{2}\left(M\right)$ is actually independent
on chart $\left(U,\boldsymbol{q}\right)$. It is easy to check complete
relation $\int_{M}\ast\left|x\right\rangle \left\langle x\right|=1$
and the trace of operator $O$ being $\text{tr}O=\int_{M}\ast\left\langle x\right|O\left|x\right\rangle $ with $\ast1=\sqrt{g}\text{d}q^{1}\land\cdots\land\text{d}q^{2n}$ being
the volume element. Thus the Eq. (\ref{eq:Witten}) reads $\textrm{ind}\cancel{D}^{+}=\int_{M}I$ with integrand
\begin{equation}
I\left(x\right)=\ast\textrm{tr}_{C\ell\otimes E_{x}}\left(-1\right)^{F}\left\langle x\right|e^{-\tau H}\left|x\right\rangle .\label{eq:integrand}
\end{equation}
Then we  need to prove  $I\left(x\right)$ is a topological characteristic.

We follow the usual coordinate representation in QM to define the
ket basis $\left|\boldsymbol{q}_{\boldsymbol{q}}\right\rangle $ (abbreviated
to $\left|\boldsymbol{q}\right\rangle $) in coordinate space $\boldsymbol{q}\left(U\right)\cong\mathbb{R}^{2n}$
such that

\begin{equation}
\left\langle \boldsymbol{q}|\boldsymbol{q}'_{\boldsymbol{q}}\right\rangle =\delta\left(\boldsymbol{q}-\boldsymbol{q}'\right).
\end{equation}
It is easy to check that $\boldsymbol{q}^{\ast}\left|\boldsymbol{q}\right\rangle =g^{1/4}\left|x\right\rangle $
meets the above requirement, where $\boldsymbol{q}^{\ast}$ is the
pull back mapping. Making use of Riemannian normal coordinate $\boldsymbol{x}:U\rightarrow\mathbb{R}^{2n}$,
we adopt moving frame as $\boldsymbol{q}=\frac{1}{\sqrt{\tau}}\boldsymbol{x}:U\rightarrow\mathbb{R}^{2n}$
in the following calculation. Its metric at point $x$ reads $\tau\delta_{\mu\nu}\text{d}q^{\mu}\otimes\text{d}q^{\nu}$
with $g\left(x\right)=\tau^{2n}$, which leads to $\left|x\right\rangle =\tau^{-n/2}\boldsymbol{q}^{\ast}\left|\boldsymbol{0}_{\boldsymbol{q}}\right\rangle $.
Thus through the pull-back mapping $\boldsymbol{q}^{\ast}I\left(\boldsymbol{q}\right)|_{\boldsymbol{q}=\boldsymbol{0}}=I\left(x\right)$,
the integrand Eq. (\ref{eq:integrand}) is expressed by coordinates,
\begin{equation}
I\left(\boldsymbol{0}\right)=\frac{1}{\tau^{n}}\ast\textrm{tr}_{C\ell\otimes E_{x}}\left(-1\right)^{F}\left\langle \boldsymbol{0}_{\boldsymbol{q}}\right|e^{-\tau H}\left|\boldsymbol{0}_{\boldsymbol{q}}\right\rangle ,\label{eq:integ_phi}
\end{equation}
in which the (scaled) Hamiltonian from Eq. (\ref{eq:Hamil}) as well
as the (scaled) connection from Eq. (\ref{eq:conn}) are given by
\begin{equation}
\frac{\sqrt{\tau}}{i}D_{\alpha}=\frac{1}{i}\frac{\partial}{\partial q^{\alpha}}-\frac{1}{4\sqrt{\tau}}\Gamma_{\alpha\mu}^{\nu}\cdot i\tau\tilde{e}^{\mu}\tilde{e}_{\nu}-i\sqrt{\tau}\omega_{\alpha},
\end{equation}
\begin{equation}
\tau H=\frac{1}{2}\sum_{\alpha}\left(\frac{\sqrt{\tau}}{i}D_{\alpha}\right)^{2}+\frac{\tau}{2}\tilde{\Omega}+\frac{i}{2}\sum_{\alpha}\sqrt{\tau}\Gamma_{\alpha\alpha}^{\nu}\cdot\frac{\sqrt{\tau}}{i}D_{\nu}+\frac{\tau\mathcal{R}}{8}.
\end{equation}
Till now, the QM on $M$ has been mapped into that on $\mathbb{R}^{2n}$.

\subsection{Reduction of Clifford algebra}

We notice the trace of Clifford algebra is given by $\textrm{tr}_{\mathbb{C}\ell}1=2^{n}$
for the zero-order algebra and $\textrm{tr}_{\mathbb{C}\ell}X=0$
for the higher-order algebra, therefore, we find $\left(-1\right)^{F}\cdot2^{-n}\textrm{tr}_{\mathbb{C}\ell}\left(-1\right)^{F}$
is the $2n$-order projector of $\mathbb{C}\ell$. Denoting $\mathcal{P}^{r}:\land\left(T_{x}^{\ast}M\right)\rightarrow\land^{r}\left(T_{x}^{\ast}M\right)$
as the $r$-form projector of the exterior algebra, one finds identity
$\varphi^{-1}\left(-1\right)^{F}\cdot2^{-n}\textrm{tr}_{\mathbb{C}\ell}\left(-1\right)^{F}=\mathcal{P}^{2n}\varphi^{-1}$.
Relating it to $\varphi^{-1}\left(-1\right)^{F}=\ast i^{n}$ (see
Eq. (\ref{eq:chir})), one simplifies Eq. (\ref{eq:integ_phi}) to
\begin{equation}
I\left(\boldsymbol{0}\right)=\left(\frac{2}{i\tau}\right)^{n}\mathcal{P}^{2n}\varphi^{-1}\textrm{tr}_{E_{x}}\left\langle \boldsymbol{0}_{\boldsymbol{q}}\right|e^{-\tau H}\left|\boldsymbol{0}_{\boldsymbol{q}}\right\rangle .
\end{equation}
Next we define a single-parameter linear isomorphism $\varphi_{\epsilon}:e^{\mu_{1}}\land\cdots\land e^{\mu_{r}}\mapsto\epsilon^{r}\tilde{e}^{\mu_{1}}\cdots\tilde{e}^{\mu_{r}}$
($\epsilon\neq0,\epsilon\in\mathbb{C}$) between Clifford algebra
$\mathbb{C}\ell$ and complexified exterior algebra $\Lambda=\mathbb{C}\otimes\land\left(T_{x}^{\ast}M\right)$.
It helps distinguish the algebraic order and leads to $\frac{1}{\epsilon^{r}}\mathcal{P}^{r}\varphi^{-1}=\mathcal{P}^{r}\varphi_{\epsilon}^{-1}$.
Choosing $\epsilon^{2}=i\pi\tau$, one finds $\tau H=\varphi_{\epsilon}H^{\text{ext}}$
and $\sqrt{\tau}D_{\alpha}=\varphi_{\epsilon}D_{\alpha}^{\text{ext}}$
with
\begin{equation}
H^{\text{ext}}=\frac{1}{2}\sum_{\alpha}\left(\frac{1}{i}D_{\alpha}^{\text{ext}}\right)^{2}+\frac{i}{2}\sum_{\alpha}\sqrt{\tau}\Gamma_{\alpha\alpha}^{\nu}\cdot\frac{1}{i}D_{\nu}^{\text{ext}}-\frac{i}{2\pi}\Omega+\frac{\tau\mathcal{R}}{8},\label{eq:Ham_ext}
\end{equation}
\begin{equation}
\frac{1}{i}D_{\alpha}^{\text{ext}}=\frac{1}{i}\frac{\partial}{\partial q^{\alpha}}-\frac{1}{4\pi\sqrt{\tau}}\Gamma_{\alpha\mu}^{\nu}\cdot e^{\mu}\land e_{\nu}-i\sqrt{\tau}\omega_{\alpha}.\label{eq:conn_ext}
\end{equation}
Thus the integrand is reduced to

\begin{equation}
I\left(\boldsymbol{0}\right)=\left(2\pi\right)^{n}\mathcal{P}^{2n}\varphi_{\epsilon}^{-1}\textrm{tr}_{E_{x}}\left\langle \boldsymbol{0}_{\boldsymbol{q}}\right|e^{-\varphi_{\epsilon}H^{\text{ext}}}\left|\boldsymbol{0}_{\boldsymbol{q}}\right\rangle .
\end{equation}
Since $\lim_{\left|\epsilon\right|\rightarrow0}\left\{ \varphi_{\epsilon}\left(e^{\mu}\right),\varphi_{\epsilon}\left(e^{\nu}\right)\right\} =0\in\mathbb{C}\ell$
displays the anti-commutation relation of exterior algebra, then $\varphi_{\epsilon}:\Lambda\rightarrow\mathbb{C}\ell$
becomes an algebraic isomorphism when $\left|\epsilon\right|\rightarrow0$,
i.e., $\lim_{\left|\epsilon\right|\rightarrow0}\varphi_{\epsilon}^{-1}\left[\varphi_{\epsilon}\left(\xi\right)\cdot\varphi_{\epsilon}\left(\eta\right)\right]=\xi\land\eta$
for $\forall\xi,\eta\in\Lambda$. Therefore, we obtain
\begin{equation}
\lim_{\tau\rightarrow0}I\left(\boldsymbol{0}\right)=\left(2\pi\right)^{n}\mathcal{P}^{2n}\lim_{\tau\rightarrow0}\textrm{tr}_{E_{x}}\left\langle \boldsymbol{0}_{\boldsymbol{q}}\right|e^{-H^{\text{ext}}}\left|\boldsymbol{0}_{\boldsymbol{q}}\right\rangle ,\label{eq:reduce}
\end{equation}
in which the Clifford algebra in the integrand has been reduced
to the differential form.

\subsection{Derivation of topological invariant}

We clearly see that the terms $\sqrt{\tau}\Gamma_{\alpha\alpha}^{\nu}$,
$\sqrt{\tau}\omega_{\alpha}$ and $\tau\mathcal{R}$ in Eq. (\ref{eq:Ham_ext})
and Eq. (\ref{eq:conn_ext}) vanish in the limit $\tau\rightarrow0$,
so that we just need to calculate the limit of $\Gamma_{\beta\mu}^{\nu}/\sqrt{\tau}$.
Returning back to the Riemannian normal coordinate $\boldsymbol{x}=\sqrt{\tau}\boldsymbol{q}$,
one finds
\begin{equation}
\lim_{\tau\rightarrow0}\frac{1}{\sqrt{\tau}}\Gamma_{\beta\mu}^{\nu}\left(\sqrt{\tau}\boldsymbol{q}\right)=q^{\alpha}\frac{\partial\Gamma_{\beta\mu}^{\nu}}{\partial x^{\alpha}}|_{\boldsymbol{x}=\boldsymbol{0}}.
\end{equation}
With the help of the definition $R_{\mu\alpha\beta}^{\nu}=\frac{\partial}{\partial x^{\alpha}}\Gamma_{\beta\mu}^{\nu}-\frac{\partial}{\partial x^{\beta}}\Gamma_{\alpha\mu}^{\nu}$
and identity $\frac{\partial}{\partial x^{\alpha}}\Gamma_{\beta\mu}^{\nu}+\frac{\partial}{\partial x^{\beta}}\Gamma_{\mu\alpha}^{\nu}+\frac{\partial}{\partial x^{\mu}}\Gamma_{\alpha\beta}^{\nu}=0$
at $\boldsymbol{x}=\boldsymbol{0}$, one obtains the expansion of
the connection 
\begin{equation}
\frac{\partial}{\partial x^{\alpha}}\Gamma_{\beta\mu}^{\nu}=-\frac{1}{3}\left(R_{\mu\beta\alpha}^{\nu}+R_{\beta\mu\alpha}^{\nu}\right).\label{eq:expan}
\end{equation}
From the first Bianchi identity as well as the symmetric properties
of the Riemann tensor, one gets $\left(2R_{\nu\beta\mu\alpha}-R_{\alpha\beta\mu\nu}\right)e^{\mu}\land e^{\nu}=0.$
Combining this equation with Eq. (\ref{eq:expan}), we thus obtain
\begin{equation}
\frac{\partial}{\partial x^{\alpha}}\Gamma_{\beta\mu}^{\nu}\cdot e^{\mu}\land e_{\nu}=-\frac{1}{2}R_{\alpha\beta\mu\nu}e^{\mu}\land e^{\nu}.
\end{equation}
Up to now, the limit $\tau\rightarrow0$ in Eq. (\ref{eq:reduce})
is completely worked out,
\begin{equation}
I\left(\boldsymbol{0}\right)=\mathcal{P}^{2n}\left(2\pi\right)^{n}\left\langle \boldsymbol{0}_{\boldsymbol{q}}\right|e^{-\frac{1}{2}\sum_{\beta=1}^{2n}\left(\frac{1}{i}\frac{\partial}{\partial q^{\beta}}+\frac{1}{8\pi}q^{\alpha}R_{\alpha\beta\mu\nu}e^{\mu}\land e^{\nu}\right)^{2}}\left|\boldsymbol{0}_{\boldsymbol{q}}\right\rangle \textrm{tr}_{E_{x}}e^{\frac{i}{2\pi}\Omega}.\label{eq:topology}
\end{equation}

To see the structure of Eq. (\ref{eq:topology}) clearly, we treat
$p_{\beta}=\frac{1}{i}\frac{\partial}{\partial q^{\beta}}$ and $q^{\alpha}$
as the momentum and spatial displacement operators in the QM on coordinate
space $\boldsymbol{q}\left(U\right)\cong\mathbb{R}^{n}$. We also
denote the eigen state of an operator $\hat{O}$ by $\left|\lambda_{\hat{O}}\right\rangle $
for eigen value $\lambda$ (with abbreviation $\left|O_{\hat{O}}\right\rangle $
to $\left|O\right\rangle $), so that $\left|\boldsymbol{0}_{\boldsymbol{q}}\right\rangle $
is denoted by $\left|0_{q^{1}}\right\rangle \otimes\cdots\otimes\left|0_{q^{2n}}\right\rangle $.
Furthermore, we apply the splitting principle of the characteristic
$\frac{1}{4\pi}R_{2l-1,2l,\mu\nu}e^{\mu}\land e^{\nu}=-y_{l}\in\Omega^{2}\left(M\right)$
($l=1,\cdots,n$). Thus Eq. (\ref{eq:topology}) is decomposed to
\begin{equation}
I\left(\boldsymbol{0}\right)=\mathcal{P}^{2n}\prod_{l=1}^{n}A_{l}\land\text{ch}\left(E\right)|_{\boldsymbol{q}=\boldsymbol{0}},\label{eq:chara}
\end{equation}
where $\textrm{tr}_{E}e^{\frac{i}{2\pi}\Omega}=\text{ch}\left(E\right)$
is the Chern character of the vector bundle $E$, and factor $A_{l}$
is given by
\begin{equation}
A_{l}=2\pi\left\langle 0_{q^{2l-1}}0_{q^{2l}}\right|e^{-\frac{1}{2}\left[\left(p_{2l}-\frac{1}{2}y_{l}q^{2l-1}\right)^{2}+\left(p_{2l-1}+\frac{1}{2}y_{l}q^{2l}\right)^{2}\right]}\left|0_{q^{2l-1}}0_{q^{2l}}\right\rangle .\label{eq:A-factor}
\end{equation}
We find that $\prod_{l=1}^{n}A_{l}$ is independent on the choice
of the unitary frame $e^{\mu}\left(x\right)$, so that the integrand
$I$ is actually an $SO\left(2n\right)$-invariant polynomial regarding
to the curvature forms. According to the Chern-Weil homomorphism,
the integrand $I$ is a cohomology class which is the characteristic
of the twisted spinor bundle. Therefore, the local Atiyah-Singer index
theorem is proved.

\subsection{Calculation of $\hat{A}$ characteristic}

The calculation of characteristic Eq. (\ref{eq:A-factor}), power
series of 2-form $y_{l}$, becomes rather straightforward. Its value
can be extracted by $A_{l}=A\left(y_{l}\right)$ with the generating
function

\begin{equation}
A\left(y\right)=2\pi\left\langle 0_{\hat{q}_{1}}0_{\hat{q}_{2}}\right|e^{-\frac{1}{2}\left[\left(\hat{p}_{2}-\frac{y}{2}\hat{q}_{1}\right)^{2}+\left(\hat{p}_{1}+\frac{y}{2}\hat{q}_{2}\right)^{2}\right]}\left|0_{\hat{q}_{1}}0_{\hat{q}_{2}}\right\rangle ,y\in\mathbb{R},\label{eq:genfun}
\end{equation}
where $\hat{q}_{1,2}$ and $\hat{p}_{1,2}$ are the usual displacement
and momentum operators for QM on $\mathbb{R}^{2}$. Since $A\left(y\right)$
is an even function, we just consider the case of $y\geq0$. After
making canonical transformation 
\begin{equation}
\hat{Q}_{1}=\frac{1}{\sqrt{y}}\hat{p}_{2}+\frac{\sqrt{y}}{2}\hat{q}_{1},\hat{P}_{1}=\frac{1}{\sqrt{y}}\hat{p}_{1}-\frac{\sqrt{y}}{2}\hat{q}_{2},\label{eq:ConTsf1}
\end{equation}
\begin{equation}
\hat{Q}_{2}=-\frac{1}{\sqrt{y}}\hat{p}_{2}+\frac{\sqrt{y}}{2}\hat{q}_{1},\hat{P}_{2}=\frac{1}{\sqrt{y}}\hat{p}_{1}+\frac{\sqrt{y}}{2}\hat{q}_{2},\label{eq:ConTsf2}
\end{equation}
with $\left[\hat{Q}_{1},\hat{P}_{1}\right]=\left[\hat{Q}_{2},\hat{P}_{2}\right]=i$,
we eliminate one degree of freedom in Eq. (\ref{eq:genfun}) and arrive
at 
\begin{equation}
A\left(y\right)=2\pi\left\langle 0_{\hat{q}_{1}}0_{\hat{q}_{2}}\right|e^{-\frac{y}{2}\left(\hat{Q}_{2}^{2}+\hat{P}_{2}^{2}\right)}\left|0_{\hat{q}_{1}}0_{\hat{q}_{2}}\right\rangle .\label{eq:trans}
\end{equation}
Here $\frac{y}{2}\left(\hat{Q}_{2}^{2}+\hat{P}_{2}^{2}\right)$ represents
the Hamiltonian of a quantum harmonic oscillator (QHO).

We also need to express $\left|0_{\hat{q}_{1}}0_{\hat{q}_{2}}\right\rangle $
by the eigen states for new canonical variables. Actually $\left|q_{1}p_{2}\right\rangle $
is the eigen state of operators $\hat{Q}_{1}$ and $\hat{Q}_{2}$
as seen in Eqs. (\ref{eq:ConTsf1}-\ref{eq:ConTsf2}), which means
$\left|q_{1}p_{2}\right\rangle =C\left|Q_{1}Q_{2}\right\rangle $
with $C\in\mathbb{C}$. After choosing the following convention of
representations,
\begin{equation}
\left\langle q|q'\right\rangle =\delta\left(q-q'\right),\left\langle p|p'\right\rangle =\delta\left(p-p'\right),\label{eq:norm}
\end{equation}
\begin{equation}
\left|p\right\rangle =\int dq\frac{1}{\sqrt{2\pi}}e^{ipq}\left|q\right\rangle ,\left|q\right\rangle =\int dp\frac{1}{\sqrt{2\pi}}e^{-ipq}\left|p\right\rangle ,\label{eq:represent}
\end{equation}
we find $C=1$ up to a phase factor. Then the relation between the
old and new eigen states is established,
\begin{eqnarray}
\left|0_{\hat{q}_{1}}0_{\hat{q}_{2}}\right\rangle  & = & \int dq_{1}\frac{dp_{2}}{\sqrt{2\pi}}\delta\left(q_{1}\right)e^{-ip_{2}0}\left|q_{1}p_{2}\right\rangle \nonumber \\
 & = & \frac{1}{\sqrt{2\pi}}\int dQ_{1}dQ_{2}\cdot\delta\left(\frac{Q_{1}+Q_{2}}{\sqrt{y}}\right)\left|Q_{1}Q_{2}\right\rangle \nonumber \\
 & = & \sqrt{\frac{y}{2\pi}}\int dQ_{2}\cdot\left|\left(-Q_{2}\right)_{\hat{Q}_{1}}\right\rangle \otimes\left|Q_{2}\right\rangle .
\end{eqnarray}
Substituting it to Eq. (\ref{eq:trans}), one obtains
\begin{equation}
A\left(y\right)=y\int dQ_{2}\left\langle Q_{2}\right|e^{-\frac{y}{2}\left(\hat{Q}_{2}^{2}+\hat{P}_{2}^{2}\right)}\left|Q_{2}\right\rangle =y\cdot\textrm{tr}e^{-\frac{y}{2}\left(\hat{Q}_{2}^{2}+\hat{P}_{2}^{2}\right)}.
\end{equation}

We see that $A\left(y\right)$ is closely related to the partition
function of QHO. Since the eigenvalue of $\frac{1}{2}\left(\hat{Q}_{2}^{2}+\hat{P}_{2}^{2}\right)$
reads $m+\frac{1}{2}$ with non-negative integer $m$, we reach 
\begin{equation}
A\left(y\right)=ye^{-y/2}\sum_{m=0}^{\infty}e^{-my}=\frac{y/2}{\sinh\left(y/2\right)}.
\end{equation}
Therefore, factor $\prod_{l=1}^{n}A_{l}=\prod_{l=1}^{n}\frac{y_{l}/2}{\sinh\left(y_{l}/2\right)}$
is exactly the $\hat{A}$ characteristic for the manifold $M$. Relating
it to Eq. (\ref{eq:chara}), we obtain $I\left(x\right)=\mathcal{P}^{2n}\hat{A}\left(TM\right)\land\text{ch}\left(E\right)$.
Finally, we make integration and prove the Atiyah-Singer index theorem
\begin{equation}
\textrm{ind}\cancel{D}^{+}=\int_{M}\hat{A}\left(TM\right)\land\text{ch}\left(E\right).
\end{equation}

\section{Summary\label{sec:Summary}}

The profound Atiyah-Singer theorem of Dirac operator have given many
inspirations to physicists and mathematician to understand the geometry,
topology and quantum behaviour of particles and fields in microscopic
world. In mathematics, it relates analysis to topology, and in physics,
it relates high temperature physics to low temperature physics. We
illustrate how this theorem can be concisely proved in the canonical
formulation of QM. Firstly, the analytical index is expressed as the
Witten index in the QM with the Hamiltonian given by the Weitzenb\"{o}ck
identity. In the derivation, the SUSY is naturally encoded in the
properties of Hodge's theory and Clifford algebra. Secondly, in the
canonical formulation of QM, the Clifford algebra is reduced to the
exterior differential form through their algebraic isomorphism in
small $\tau$ limit, leading to the emergence of topological characteristics.
This step plays a similar role as the Localization technique in the
path integral formulation for the SUSY-QM models. Finally, the expression
of $\hat{A}$ characteristic is obtained with the help of the property
of quantum harmonic oscillator, while the Chern character naturally
appears in the topological index. Therefore, our proof can be regarded
as an independent version to the SUSY-QM proof by using path integral.

\end{document}